# Quantum Communication Network Utilizing Quadripartite Entangled States of Optical Field

Heng Shen, Xiaolong Su,* Xiaojun Jia, and Changde Xie

State Key Laboratory of Quantum Optics and Quantum Optics Devices, Institute of Opto-Electronics,
Shanxi University, Taiyuan 030006, People's Republic of China

**Abstract**: We propose two types of quantum dense coding communication networks with optical continuous variables, in which a quadripartite entangled state of the optical field with totally three-party correlations of quadrature amplitudes is utilized. In the networks, the exchange of information between any two participants can be manipulated by one or two of the remaining participants. The channel capacities for a variety of communication protocols are numerically calculated. Due to the fact that the quadripartite entangled states applied in the communication systems have been successfully prepared already in the laboratory, the proposed schemes are experimentally accessible at present.

**PACS number(s)**: 03.67.Hk, 42.50.Dv

## I. Introduction

Since quantum teleportation was proposed by Bennett *et. al* in 1993 [1], theoretical and experimental researches on quantum information have developed rapidly. As one of the important applications of quantum entanglement, quantum dense coding, which is a way to transmit two bits of information through the manipulation of only one of two entangled quantum systems [2], has been experimentally demonstrated in both discrete and continuous quantum variable regimes [3-7]. Successively, controlled dense coding for continuous variables (CV) utilizing tripartite Greenberger-Horne-Zeilinger (GHZ) states of light has also been experimentally realized following the theoretical proposals [8-10]. The investigation of the classical information capacities of noisy quantum communication channels is a significant component of contemporary quantum information theory. It has been well-demonstrated that the classical capacity of some noisy quantum channels can be increased to a higher level relative to the best known classical capacity achievable without entanglement, which is named the entanglement-assisted classical capacity [11, 12]. Applying multipartite entanglement and dense coding in quantum communication networks we can realize a variety of entanglement-assisted network communications with channel capacities higher than their classical limits. Recently, CV four-partite entangled states of the optical field with total three-party correlations (TTPC) have been successfully prepared in the laboratory [13]. Based on the use of the TTPC state, we designed different protocols for controlled dense coding communication.

In this paper, we propose two types of quantum communication networks in which the four subsystems of a TTPC entangled state off-line prepared are distributed to four distant communication stations and different schemes of controlled dense coding communication can be realized. Exploiting the special CV quantum multipartite entanglement, balanced homodyne detections and classical communication, the information exchange carried by both amplitude and phase quadratures of optical fields between any two participants can be manipulated by one or two of the remaining participants.

## II. TTPC entanglement

It has been theoretically and experimentally demonstrated that if two subsystems ($\hat{a}_2$ and $\hat{a}_3$ in Fig.1) from two pairs ($\hat{a}_1$, $\hat{a}_2$ and $\hat{a}_3$, $\hat{a}_4$) of independent Einstein-Podolosy-Rosen (EPR) entangled optical beams are combined on a 50% beamsplitter (BS) with a relative phase of π/2 (FS), the output two optical modes, $\hat{b}_2$ and $\hat{b}_3$, and the retained two modes, $\hat{a}_1$ and $\hat{a}_4$, constitute a quadripartite TTPC entangled state [13]. The four modes can be expressed by

$$\hat{b}_1 = \hat{a}_1 = \hat{X}_{a_1} + i\hat{Y}_{a_1},$$

$$\hat{b}_2 = \frac{1}{\sqrt{2}}(\hat{a}_2 + i\hat{a}_3) = \frac{1}{\sqrt{2}}(\hat{X}_{a_2} - \hat{Y}_{a_3}) + \frac{i}{\sqrt{2}}(\hat{Y}_{a_2} + \hat{X}_{a_3}),$$

$$\hat{b}_3 = \hat{a}_4 = \hat{X}_{a_4} + i\hat{Y}_{a_4},$$

$$\hat{b}_4 = \frac{1}{\sqrt{2}}(\hat{a}_2 - i\hat{a}_3) = \frac{1}{\sqrt{2}}(\hat{X}_{a_2} + \hat{Y}_{a_3}) + \frac{i}{\sqrt{2}}(\hat{Y}_{a_2} - \hat{X}_{a_3}), \quad (1)$$

where $\hat{X}_{a_i}$ and $\hat{Y}_{a_i}$ ($i$ = 1, 2, 3, 4) are the amplitude and phase quadratures of modes $\hat{a}_i$ respectively. Assuming the two EPR pairs are produced from two independent nondegenerate optical parametric amplifiers (NOPAs) operating at deamplification, i.e. the phase difference between the pump light and seed light is π, the amplitude and phase quadratures of the output modes $a_i$ are expressed by [15]

$$\hat{X}_{a_1(a_3)} = \hat{X}_{01(03)} \cosh r - \hat{X}_{02(04)} \sinh r,$$

$$\hat{Y}_{a_1(a_3)} = \hat{Y}_{01(03)} \cosh r + \hat{Y}_{02(04)} \sinh r,$$

$$\hat{X}_{a_2(a_4)} = \hat{X}_{02(04)} \cosh r - \hat{X}_{01(03)} \sinh r,$$

$$\hat{Y}_{a_2(a_4)} = \hat{Y}_{02(04)} \cosh r + \hat{Y}_{01(03)} \sinh r, \quad (2)$$

where $X_{0i}$ and $Y_{0i}$ ($i$ = 1, 2, 3, 4) are the initial amplitude and phase quadratures of the signal light injected into the NOPA, which are the optical beams in the coherent state [9,10]. For simplification and without loss of generality, we have assumed that the two EPR entangled states have the same quantum anticorrelations of amplitude quadratures and the correlation of phase quadratures, that is [13]

$$\left\langle \delta^2(\hat{X}_{a1(a3)} + \hat{X}_{a2(a4)}) \right\rangle = \left\langle \delta^2(\hat{Y}_{a1(a3)} - \hat{Y}_{a2(a4)}) \right\rangle = 2e^{-2r}, \tag{3}$$

where $0 < r < \infty$ is the correlation parameter of the EPR entangled states, $r = 0$ without any quantum correlation and $r \rightarrow \infty$ corresponding to the ideal correlation [15]. The quadrature variances of the injected signals in the coherent state are normalized $V(\hat{X}_{0i}) = V(\hat{Y}_{0i}) = 1$. In the ideal case of $r \rightarrow \infty$, from Eq. (1) and Eq. (3) we obtain

$$\begin{aligned}
&\text{I} \quad \sqrt{2}\hat{X}_{b_1} + \hat{X}_{b_2} + \hat{X}_{b_4} = \sqrt{2}(\hat{X}_{01} + \hat{X}_{02})e^{-r} \rightarrow 0, \\
&\text{II} \quad \hat{Y}_{b_2} + \sqrt{2}\hat{X}_{b_3} - \hat{Y}_{b_4} = \sqrt{2}(\hat{X}_{03} + \hat{X}_{04})e^{-r} \rightarrow 0, \\
&\text{III} \quad -\sqrt{2}\hat{Y}_{b_1} + \hat{Y}_{b_2} + \hat{Y}_{b_4} = -\sqrt{2}(\hat{Y}_{01} - \hat{Y}_{02})e^{-r} \rightarrow 0, \\
&\text{IV} \quad \hat{X}_{b_2} + \sqrt{2}\hat{Y}_{b_3} - \hat{X}_{b_4} = -\sqrt{2}(\hat{Y}_{03} - \hat{Y}_{04})e^{-r} \rightarrow 0, \\
&\text{V} \quad \hat{Y}_{b_1} + \hat{X}_{b_3} - \sqrt{2}\hat{Y}_{b_4} = [(\hat{X}_{03} + \hat{X}_{04}) + (\hat{Y}_{01} - \hat{Y}_{02})]e^{-r} \rightarrow 0, \\
&\text{VI} \quad -\hat{Y}_{b_1} + \sqrt{2}\hat{Y}_{b_2} + \hat{X}_{b_3} = [(\hat{X}_{03} + \hat{X}_{04}) - (\hat{Y}_{01} - \hat{Y}_{02})]e^{-r} \rightarrow 0, \\
&\text{VII} \quad \hat{X}_{b_1} - \hat{Y}_{b_3} + \sqrt{2}\hat{X}_{b_4} = [(\hat{X}_{01} + \hat{X}_{02}) + (\hat{Y}_{03} - \hat{Y}_{04})]e^{-r} \rightarrow 0, \\
&\text{VIII} \quad \hat{X}_{b_1} + \sqrt{2}\hat{X}_{b_2} + \hat{Y}_{b_3} = [(\hat{X}_{01} + \hat{X}_{02}) - (\hat{Y}_{03} - \hat{Y}_{04})]e^{-r} \rightarrow 0,
\end{aligned} \tag{4}$$

where $\hat{X}_{b_i}$ and $\hat{Y}_{b_i}$ ($i = 1, 2, 3, 4$) are the amplitude and the phase quadratures of modes $\hat{b}_i$, respectively. In all these correlations the three-part combinations of quadratures from the four submodes are involved. The correlation properties are different from that of CV cluster and GHZ quadripartite entangled states, in which some two-part correlations are included [16, 17]. We can see that there are two types of correlations in these three-party correlations. The correlations I-IV depend on two quadratures of the seed light only and the correlations V-VIII depend on four quadratures of them. The correlation relations of TTPC entangled state can be denoted by the Graph G (representing a type of weighted graph states [14]) in Fig. 1 (b), where the vertices denote the optical modes, the edges (line that connected two neighbor vertices) represent the interaction between the optical modes. The amplitude and phase quadratures of every three parts of the TTPC state are correlated according to the relations given by Eq. (4). If the four submodes are distributed to four stations, the communication networks including four users can be constructed.

## III. Dense coding communication networks consisting of four stations sharing TTPC state

We analyze a quantum communication network consisting of four participants Alice, Bob, Claire and Daisy (Fig.2). The four entangled submodes $\hat{b}_1, \hat{b}_2, \hat{b}_3$ and $\hat{b}_4$ from a TTPC source are distributed to the four distant stations. We will demonstrate that the dense coding quantum communication [6] can be realized between any two stations in the network only under the help of one or two other participants. Due to applying the TTPC entanglement the information amounts transmitted by both amplitude and phase quadratures of a submode between any two users can be controlled by other participants.

**A. Communication between two users without the help of others**

For comparison, we will discuss the communication between Alice and Bob (Claire) without the help of other users. Since there are complex quantum correlations [see Eq. (4)], the communication protocols between two connected neighbors (Alice and Bob, Alice and Daisy, Bob and Claire, Claire and Daisy) and that between two users on the diagonal lines (Alice and Claire, Bob and Daisy) are different [see Fig. 1 (b)]. For example, we will analyze the communication between Alice and Bob as well as Alice and Claire, respectively. For the communication between two neighbors, such as Alice and Bob, Alice modulates the classical amplitude ($X_s$) and phase signals ($Y_s$) respectively on the amplitude and phase quadratures of her mode $\hat{b}_1$ by means of amplitude and phase modulators, to make $\hat{b}_1' = \hat{b}_1 + a_s$ ($a_s = X_s + iY_s$) and then sends $\hat{b}_1'$ to Bob, who demodulates the modulated information with a Bell-state direct detection system [18], under the help of his submode $\hat{b}_2$. From Eq. (1), we can calculate the sum and difference photocurrents detected by Bob

$$\begin{aligned}
i_+ &= \frac{1}{\sqrt{2}}(\hat{X}_{b_2} + \hat{X}_{b_1'}) \\
&= \frac{1}{2}[(\sqrt{2}\cosh r - \sinh r)\hat{X}_{01} + (\cosh r - \sqrt{2}\sinh r)\hat{X}_{02} - \cosh r \hat{Y}_{03} - \sin r \hat{Y}_{04}] + \frac{X_s}{\sqrt{2}}, \\
i_- &= \frac{1}{\sqrt{2}}(\hat{Y}_{b_2} - \hat{Y}_{b_1'}) \\
&= \frac{1}{2}[(\cosh r - \sqrt{2}\sinh r)\hat{Y}_{02} + (\sinh r - \sqrt{2}\cosh r)\hat{Y}_{01} + \cosh r \hat{X}_{03} - \sin r \hat{X}_{04}] - \frac{Y_s}{\sqrt{2}}.
\end{aligned} \quad (5)$$

The transmitted amplitude ($X_s$) and phase ($Y_s$) signals are involved in $i_+$ and $i_-$, respectively. The relevant power spectra of the photocurrents are

$$\begin{aligned}
\langle \delta^2(i_+) \rangle_{AB} &= \frac{[(\sqrt{2}-1)^2+1]e^{2r} + [(\sqrt{2}+1)^2+1]e^{-2r}}{8} + \frac{V_{X_s}}{2} \\
&= \frac{(4-2\sqrt{2}) + (4+2\sqrt{2})s^2}{8s} + \frac{V_{X_s}}{2}, \\
\langle \delta^2(i_-) \rangle_{AB} &= \frac{[(\sqrt{2}-1)^2+1]e^{2r} + [(\sqrt{2}+1)^2+1]e^{-2r}}{8} + \frac{V_{Y_s}}{2} \\
&= \frac{(4-2\sqrt{2}) + (4+2\sqrt{2})s^2}{8s} + \frac{V_{Y_s}}{2},
\end{aligned} \quad (6)$$

where $s = e^{-2r}$ represents the squeezing degree of the output light field from the NOPAs.

Similarly, for the communication between Alice and Claire on the diagonal line without the help of other users, the calculated power spectra of sum and difference photocurrents are

$$\begin{aligned}
\langle \delta^2(i_+) \rangle_{AC} &= \frac{e^{-2r} + e^{2r}}{2} + \frac{1}{2}V_{X_s}, \\
\langle \delta^2(i_-) \rangle_{AC} &= \frac{e^{-2r} + e^{2r}}{2} + \frac{1}{2}V_{Y_s}.
\end{aligned} \quad (7)$$

We can see from Eq. (6) and Eq. (7), even if $r \to \infty$, neither the amplitude signal nor the phase signal can be acquired with high accuracy due to the existence of the antisqueezing terms ($e^{2r}$). It means that for realizing efficient quantum communication between any two users in the network we have to ask the help of other users. In the following, we will discuss two types of quantum communication networks: communication between two users under the control of other two users (two controllers), and communication between two users under the control of one of the other two users (one controller).

**B. Communication with two controllers**

**B1. Communication between two neighbor users under the control of other two users**

Fig. 2 shows the communication between Alice and Bob under the control of Claire and Daisy. In this scheme, the correlation relations I and VI in Eq. (4) are utilized. Daisy and Claire directly detect the amplitude signals ($\hat{X}_{b_4}$ and $\hat{X}_{b_3}$) of their modes with balanced homodyne detectors BHD$_d$ and BHD$_c$ respectively, and send the detected $\hat{X}_{b_4}$ and $\hat{X}_{b_3}$ to Bob. Alice sends the signal beam $\hat{b}_1'$ to Bob, where the beam $\hat{b}_1'$ and Bob's beam $\hat{b}_2$ are mixed on a 1:2 beam splitter (BS1) with zero phase difference. The quadratures of two output beams from BS1 are detected by two sets of Bell-state direct detectors (BHD$_1$, BHD$_2$), respectively [18]. The detected photocurrents stand for the amplitude sum ($i_+$) and the phase difference ($i_-$) of the two input optical modes, $\hat{b}_1'$ and $\hat{b}_2$, respectively [18]. Then Bob adding $\hat{X}_{b_4}$ to $i_+$ and $\hat{X}_{b_3}$ to $i_-$ respectively (Fig. 2), the resultant photocurrents are

$$\begin{aligned}
i^{C+D}_{+AB} &= \frac{1}{\sqrt{3}}(\sqrt{2}\hat{X}_{b_1'} + \hat{X}_{b_2}) + g_x \hat{X}_{b_4} \\
&= \frac{1}{\sqrt{6}}[(2\cosh r - \sinh r - \sqrt{3}g_x \sin r)\hat{X}_{01} + (\cosh r - 2\sinh r + \sqrt{3}g_x \cosh r)\hat{X}_{02} \\
&\quad + (\sqrt{3}g_x - 1)\cosh r \hat{Y}_{03} + (\sqrt{3}g_x - 1)\sinh r \hat{Y}_{04}] + \sqrt{\frac{2}{3}} X_s, \\
i^{C+D}_{-AB} &= \frac{1}{\sqrt{3}}(\sqrt{2}\hat{Y}_{b_2} - \hat{Y}_{b_1'}) + g_y \hat{X}_{b_3} \\
&= \frac{1}{\sqrt{3}}[(\hat{Y}_{02} - \hat{Y}_{01})e^{-r} + (\cosh r - \sqrt{3}g_y \sinh r)\hat{X}_{03} + (\sqrt{3}g_y \cosh r - \sinh r)\hat{X}_{04}] - \frac{1}{\sqrt{3}} Y_s.
\end{aligned} \quad (8)$$

The corresponding power spectra are expressed by

$$\langle \delta^2(i_+) \rangle_{AB}^{C+D} = \frac{1}{12}\{2(1-\sqrt{3}g_x)^2 e^{2r} + [(3+\sqrt{3}g_x)^2 + (\sqrt{3}g_x-1)^2]e^{-2r}\} + \frac{2}{3}V_{X_s},$$

$$\langle \delta^2(i_-) \rangle_{AB}^{C+D} = \frac{1}{6}\{(\sqrt{3}g_y-1)^2 e^{2r} + [4+(\sqrt{3}g_y+1)^2]e^{-2r}\} + \frac{1}{3}V_{Y_s}, \quad (9)$$

where $g_x$ and $g_y$ are the gains at Bob for the transformation from Daisy's and Claire's photocurrents to his sum and difference photocurrents, respectively. When $g_x = g_y = 1/\sqrt{3}$ the power spectra of the sum and difference photocurrents can be denoted as

$$\langle \delta^2(i_+) \rangle_{AB}^{C+D} = \frac{4}{3}e^{-2r} + \frac{2}{3}V_{X_s},$$

$$\langle \delta^2(i_-) \rangle_{AB}^{C+D} = \frac{4}{3}e^{-2r} + \frac{1}{3}V_{Y_s}. \quad (10)$$

It means that Bob is able to obtain the complete amplitude signal with Daisy's help and the phase signal with Claire's help if $r \to \infty$. However, in an experiment, we are not able to get perfect entanglement and thus have to consider the case with a finite $r$. To a given finite $r$ the optimum gain factors ($g_x^{opt}$ and $g_y^{opt}$) can be obtained by calculating the minimums of $\langle \delta^2(i_+) \rangle_{AB}^{C+D}$ and $\langle \delta^2(i_-) \rangle_{AB}^{C+D}$ from Eq. (9),

$$g_x^{opt} = g_y^{opt} = \frac{1}{\sqrt{3}} \frac{e^{2r} - e^{-2r}}{e^{2r} + e^{-2r}}. \quad (11)$$

Substituting $g_x^{opt}$ and $g_y^{opt}$ into Eq. (9) we have

$$\langle \delta^2(i_+) \rangle_{AB}^{C+D} = \frac{2e^{-2r}}{3} \frac{e^{-4r}+2}{e^{-4r}+1} + \frac{2}{3}V_{X_s},$$

$$\langle \delta^2(i_-) \rangle_{AB}^{C+D} = \frac{2e^{-2r}}{3} \frac{e^{-4r}+2}{e^{-4r}+1} + \frac{1}{3}V_{Y_s}. \quad (12)$$

Comparing Eq.(6) and Eq.(12) we can see that the signal-to-noise ratios (SNR) of the amplitude and phase signals transmitted from Alice to Bob under the help of Daisy and Claire are higher than that without their helps, since the influence of antisqueezing terms ($e^{2r}$) is suppressed in Eq. (12). In this way the communication between any two neighbor users can be controlled by other two users and the signal modulated on both amplitude and phase quadratures can be extracted by the receiver with sensitivity beyond the shot noise limit (SNL) [6] due to applying the TTPC entanglement distributed among the four stations.

**B2. Communication between two diagonal participants under the control of other two users**

Now, we discuss the communication between two users on a diagonal line, Alice and Claire, under the control of other two users, Bob and Daisy. The construction of the communication system is analogous with Fig.2, where we only need to exchange the positions of the receiver (Bob) and controller (Claire) and replace the 1:2 BS1 with a 1:1 beamsplitter and interfere optical beam $\hat{b}_1^{'}$ and the submode $\hat{b}_3$ on BS1 with a phase difference of π/2. Bob detects the phase quadrature of optical mode $\hat{b}_2$ and Daisy detects the amplitude quadrature of $\hat{b}_4$ by own homodyne detection system, respectively, and they send the measurement results to Claire. Applying the correlation relations VI and VII in Eq. (4) and a calculation procedure similar to that used in Section B1 we obtain the optimum power spectra of the sum and the difference photocurrents in the communication from Alice to Claire:

$$\left\langle \delta^2(i_+) \right\rangle_{AC}^{B+D} = \frac{2e^{-2r}}{e^{-4r}+1} + \frac{1}{2}V_{X_s},$$

$$\left\langle \delta^2(i_-) \right\rangle_{AC}^{B+D} = \frac{2e^{-2r}}{e^{-4r}+1} + \frac{1}{2}V_{Y_s}. \tag{13}$$

The corresponding optimum gain factors are

$$g_{x(AC)}^{opt} = g_{y(AC)}^{opt} = \frac{e^{2r}-e^{-2r}}{e^{2r}+e^{-2r}}. \tag{14}$$

**C. Communication with one controller**

To achieve the controlled dense coding communication between two users under the help of only a <u>single</u> controller, we have to implement the direct optical coupling between the controller's submode and the receiver's submode instead of the electronic feedback used in scheme B. We discuss the protocol also in two cases for the neighbor users and the diagonal users.

**C1. Communication between two neighbor users under the control of a single user**

Fig. 3 shows the schematic of the communication between two neighbor users (Alice and Bob) under the control of only one other user (Daisy). Alice and Daisy respectively send the signal beam $\hat{b}_1'$ and the submode $\hat{b}_4$ to Bob's station, where Bob couples his mode $\hat{b}_2$ and Daisy's $\hat{b}_4$ on a 50% beam splitter (BS1) with the same phase to produce the output mode $\hat{b}_{out} = \frac{1}{\sqrt{2}}(\hat{b}_2 + \hat{b}_4)$, then he implements the combining Bell-state measurement of the modes $\hat{b}_1'$ and $\hat{b}_{out}$. The two output beams from the 50% beamsplitter (BS2) are detected by the photodiodes $D_1$ and $D_2$, respectively. The sum ($i_{+D}'$) and the difference ($i_{-D}'$) of the photocurrents detected by $D_1$ and $D_2$ equal to

$$i_{+AB}^D = \frac{1}{\sqrt{2}}[\frac{1}{\sqrt{2}}(\hat{X}_{b_2} + \hat{X}_{b_4}) + \hat{X}_{b_1'}]$$

$$= \frac{1}{\sqrt{2}}[(\hat{X}_{01} + \hat{X}_{02})e^{-r} + X_s],$$

$$i_{-AB}^D = \frac{1}{\sqrt{2}}[\frac{1}{\sqrt{2}}(\hat{Y}_{b_2} + \hat{Y}_{b_4}) - \hat{Y}_{b_1'}]$$

$$= \frac{1}{\sqrt{2}}[(\hat{Y}_{02} - \hat{Y}_{01})e^{-r} - \hat{Y}_s]. \quad (15)$$

The corresponding power spectra respectively are

$$\langle \delta^2(i_+) \rangle_{AB}^D = e^{-2r} + \frac{V_{X_s}}{2},$$

$$\langle \delta^2(i_-) \rangle_{AB}^D = e^{-2r} + \frac{V_{Y_s}}{2}. \quad (16)$$

**C2. Communication between two diagonal users under control of a single user**

In order to realize the information transmission between two users on a diagonal line, such as from Alice to Claire with the help of Daisy, Claire couples Alice's mode $\hat{b}_1'$ and Claire's mode $\hat{b}_3$ on a 50% beamsplitter. However the phase difference of $\hat{b}'_1$ and $\hat{b}_3$ on the beamsplitter should be $\pi/2$ but not zero as that for two neighbors. Through a similar calculation as that used in C1 and applying the correlation relations V and VII in Eq. (4), we can obtain the sum and the difference photocurrent power spectra of the communication:

$$\left\langle \delta^2(i_+) \right\rangle^D_{AC} = e^{-2r} + \frac{V_{X_s}}{4},$$

$$\left\langle \delta^2(i_-) \right\rangle^D_{AC} = e^{-2r} + \frac{V_{Y_s}}{4}. \tag{17}$$

Comparing with Eq. (16), the intensity of the modulated signals is reduced by half, but the noise background remains unchanged.

The noise levels of amplitude and phase quadratures for above-mentioned communication systems as a function of the squeezing ($-10 \cdot \log(s)$) are shown in Fig. 4 (a) and (b), where (a) for the communication between Alice and Bob and (b) for the communication between Alice and Claire, the curve 1 is the shot-noise limit (SNL), curves 2, 3, 4 are the respective noise levels for different communication protocols between two users without any help (curve 2 for $\left\langle \delta^2(i_+) \right\rangle_{AB}$ in (a) and $\left\langle \delta^2(i_+) \right\rangle_{AC}$ in (b)), with the help of two users (curve 3 for $\left\langle \delta^2(i_+) \right\rangle^{C+D}_{AB}$ in (a) and $\left\langle \delta^2(i_+) \right\rangle^{B+D}_{AC}$ in (b)), and with the help of one user (curve 4 for $\left\langle \delta^2(i_+) \right\rangle^D_{AB}$ in (a) and $\left\langle \delta^2(i_+) \right\rangle^D_{AC}$ in (b)), respectively. From equations (6), (11), (12), (14), (16), and (17) it is obvious that the noise levels of the amplitude quadratures equal to that of the phase quadratures in the communication schemes. More important, the noise levels of the amplitude and phase signals with controllers' help are below the SNL when $r > 0$. However, without their help the noise level of the amplitude and phase noises is below the SNL only when $s < 0.16$ (7.96dB) for the communication between Alice and Bob and it will never be below the SNL for the communication between Alice and Claire. The results demonstrate the significant role of quantum entanglement in the dense coding communication networks.

## IV. Comparison of channel capacities in different communication systems

In the communication of classical information with a quantum channel, the classical capacity is an important parameter. The channel that maps Gaussian input states into Gaussian output states is a Gaussian channel [12]. General lossy channel can be regarded as a Gaussian channel, and thus the channels discussed in quantum communication are usually limited to Gaussian channels. We will use the formula for Gaussian channel to discuss the channel capacities of the proposed systems in the following. If the original signal has the Gaussian distribution [19]

$$P_\alpha = \frac{1}{\pi\sigma^2}\exp(-\frac{|\alpha|^2}{\sigma^2}), \tag{18}$$

we can use the classical channel capacity formula over Gaussian channel [20]

$$I = \frac{1}{2}\ln(1+SNR), \tag{19}$$

to calculate the mutual information between two users. In Eqs. (18) and (19), $|\alpha|^2$ is the intensity of the modulated signals, $\sigma^2$ is the average value of the signal photon numbers; and SNR stands for signal-to-noise ratio. Since the classical informations modulated on the amplitude and phase quadratures of optical field are simultaneously extracted in the dense coding communication, the concrete mutual information between two users is given by

$$I = \frac{1}{2}\ln(1+SNR_X) + \frac{1}{2}\ln(1+SNR_Y), \tag{20}$$

where $SNR_X$ and $SNR_Y$ stand for the SNR of amplitude and phase quadratures, respectively. In the variances expressions of Eqs. (6), (12), and (16) [Eqs.(7), (13) and (17)], the terms containing $V_{X_s}$ ($V_{Y_s}$) represent the variances of modulated signals, while other terms stand for the fluctuation variances of quadrature components without signals. According to these variance expressions, we can calculate the SNR for the three presented communication systems, respectively. For example, from Eq. (6) the $(SNR_X)_{AB}$ and $(SNR_Y)_{AB}$ are the ratio between variances of modulated signals ($\frac{V_{X_s}}{2}$ and $\frac{V_{Y_s}}{2}$) and the fluctuation variances of quadrature components without signals ($\frac{(4-2\sqrt{2})+(4+2\sqrt{2})s^2}{8s}$ and $\frac{(4-2\sqrt{2})+(4+2\sqrt{2})s^2}{8s}$), that are,

$$(SNR_X)_{AB} = (SNR_Y)_{AB} = \frac{8s\sigma^2}{(4-2\sqrt{2})+(4+2\sqrt{2})s^2}, \quad (21)$$

where we have supposed $\frac{V_{X_s}}{2} = \frac{V_{Y_s}}{2} = \sigma^2$ without loss of generality. Similarly, the SNR for other communication systems can be calculated based on Eqs. (7), (12), (13), (16) and (17), respectively. Substituting the calculated SNR into Eq. (20), we can obtain the mutual information for the three presented communication systems, respectively,

$$I_{AB} = \ln[1+\frac{8s\sigma^2}{(4-2\sqrt{2})+(4+2\sqrt{2})s^2}],$$

$$I_{AB}^{C+D} = \frac{1}{2}\ln[1+\frac{2\sigma^2(1+s^2)}{2s+s^3}]+\frac{1}{2}\ln[1+\frac{\sigma^2(1+s^2)}{2s+s^3}],$$

$$I_{AB}^{D} = \ln(1+\frac{\sigma^2}{s}),$$

$$I_{AC} = \ln[1+\frac{2s\sigma^2}{1+s^2}],$$

$$I_{AC}^{B+D} = \ln[1+\frac{\sigma^2(1+s^2)}{2s}],$$

$$I_{AC}^{D} = \ln[1+\frac{\sigma^2}{2s}], \quad (22)$$

where $I_{AB}$ ($I_{AC}$) is the channel capacity in the network without the helps of other participants and $I_{AB}^{D}$ ($I_{AC}^{D}$) is that with Daisy's help, while $I_{AB}^{C+D}$ ($I_{AC}^{B+D}$) is that with the help of both Claire and Daisy (Bob and Daisy).

Supposing the average photon numbers $\bar{n}$ per mode supplied to the communication systems are shared by the modulated signals and squeezing, we have

$$\bar{n} = \sigma^2 + \sinh^2 r, \quad (23)$$

when $\bar{n} = e^r \sinh r$, i.e. $\sigma^2 = \sinh r \cosh r$, we obtain the channel capacities corresponding to Eq. (22) as a function of $\bar{n}$ [19]

$$C_{AB} = \ln\{1 + \frac{8(\bar{n} + \bar{n}^2)}{[(\sqrt{2}-1)^2 + 1](2\bar{n}+1)^2 + [(\sqrt{2}+1)^2 + 1]}\},$$

$$C_{AB}^{C+D} = \frac{1}{2}\ln\{1 + (\bar{n} + \bar{n}^2) + \frac{(\bar{n} + \bar{n}^2)}{1 + 2(2\bar{n}+1)^2}\} + \frac{1}{2}\ln\{1 + \frac{\bar{n} + \bar{n}^2}{2} + \frac{1}{2}\frac{\bar{n} + \bar{n}^2}{2(2\bar{n}+1)^2 + 1}\},$$

$$C_{AB}^{D} = \ln[1 + \bar{n} + \bar{n}^2],$$

$$C_{AC} = \ln\{1 + \frac{2(\bar{n} + \bar{n}^2)}{[(2\bar{n}+1)^2 + 1]}\},$$

$$C_{AC}^{B+D} = \ln\{1 + \frac{(\bar{n} + \bar{n}^2)}{2}[1 + \frac{1}{(2\bar{n}+1)^2}]\},$$

$$C_{AC}^{D} = \ln\{1 + \frac{(\bar{n} + \bar{n}^2)}{2}\}, \tag{24}$$

Fig. 5 shows the channel capacities for the different quantum channels as a function of the average photon numbers, where (a) is the channel capacity for the communication between Alice and Bob, and (b) is that for the communication between Alice and Claire. In Fig. 5 (a) the curves 1, 2 and 3 stand for $C_{AB}$, $C_{AB}^{C+D}$ and $C_{AB}^{D}$, respectively. It can be seen that the channel capacity $C_{AB}^{D}$ is highest for a given $\bar{n}$. It means that using the communication scheme C1 in section III (Fig. 3) we can achieve a channel capacity higher than that of scheme B1 with the help of both Claire and Daisy. In Fig. 5 (b) the curves 1, 2 and 3 stand for $C_{AC}$, $C_{AC}^{B+D}$ and $C_{AC}^{D}$, respectively. We can see that $C_{AC}^{B+D}$ and $C_{AC}^{D}$ are almost equal to each other. It means that for the communication between the diagonal participants the channel capacities with the help of other two participants and that under the help of only one are almost the same. In the three types of communication systems, different correlation features and different coupling schemes are used, thus the resultant channel capacities also are various.

## V. Conclusion

We presented three types of quantum communication networks in which the TTPC entangled states of optical field are exploited. Due to the existence of TTPC quantum correlations the signal transmission between two participants can be controlled by others who share the quantum entanglement with the communication participants. Owing to the asymmetric distribution of the quantum correlations among the four submodes of the TTPC entangled state, the communication schemes and results between two users on a neighbor line and a diagonal line are not totally the same. However, in all designed systems the channel capacities can be controlled and the signal-to-noise ratios can surpass the SNL of the classical optical communication. The flexibility and the versatility of the presented communication systems are specially convenient for practical applications.

We thank Jing Zhang for the helpful discussion. This research was supported by the NSFC (Grants No. 60736040 and 10804065), NSFC Project for Excellent Research Team (Grant No. 60821004), National Basic Research Program of China (Grant No. 2006CB921101).

* Corresponding author: suxl@sxu.edu.cn

**Figure captions**

Fig. 1 (Color online) TTPC entangled state. (a): Generation system of TTPC state from two EPR entangled states, (b): TTPC state in Graph picture. BS: 50% optical beam splitter, PS: phase shifter.

Fig. 2 The communication between Alice and Bob under the control of Claire and Daisy. Local: local oscillated light, BS: beam splitter, AM: amplitude modulator, PM: phase modulator, BHD: balance homodyne detector, PS: phase shifter.

Fig. 3 The information exchange process between Alice and Bob with the help of Daisy.

Fig. 4 Noise levels of phase and amplitude quadratures as a function of the squeezing degree. (a) for the communication between Alice and Bob, and (b) for the communication between Alice and Claire.

Fig. 5 The channel capacity as a function of the average photon numbers ($\bar{n}$). (a): channel capacity between Alice and Bob, (b): channel capacity between Alice and Claire.

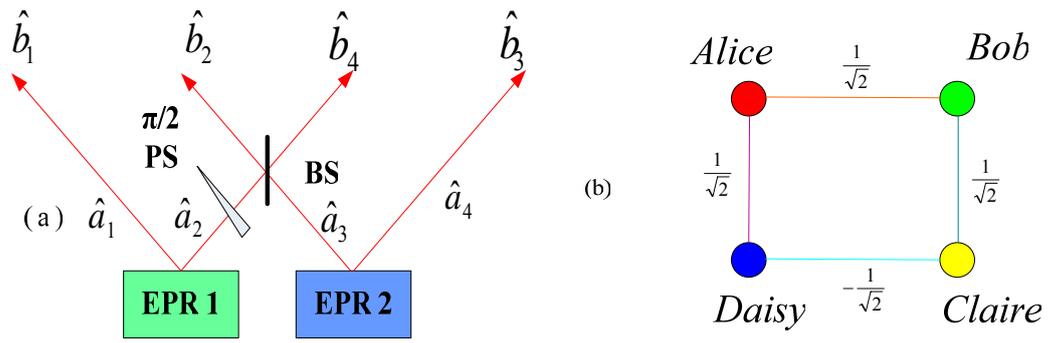

Fig. 1

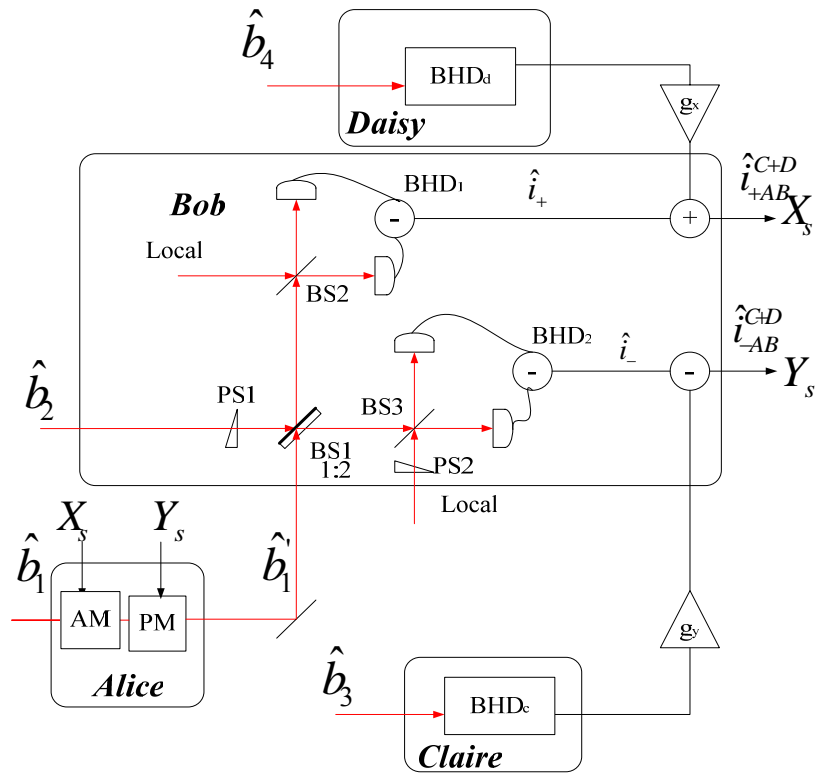

Fig. 2

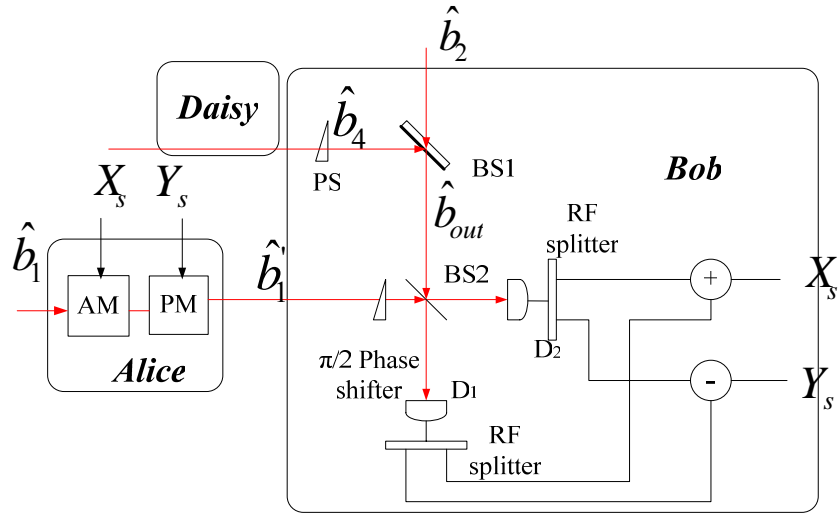

Fig. 3

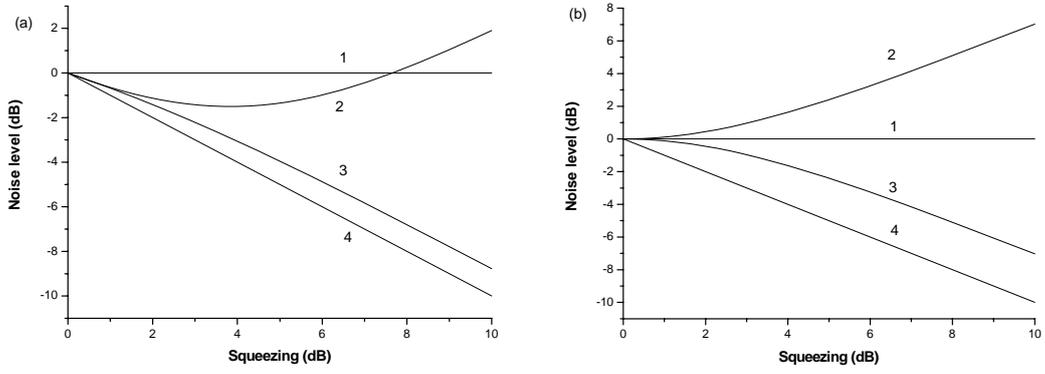

Fig. 4

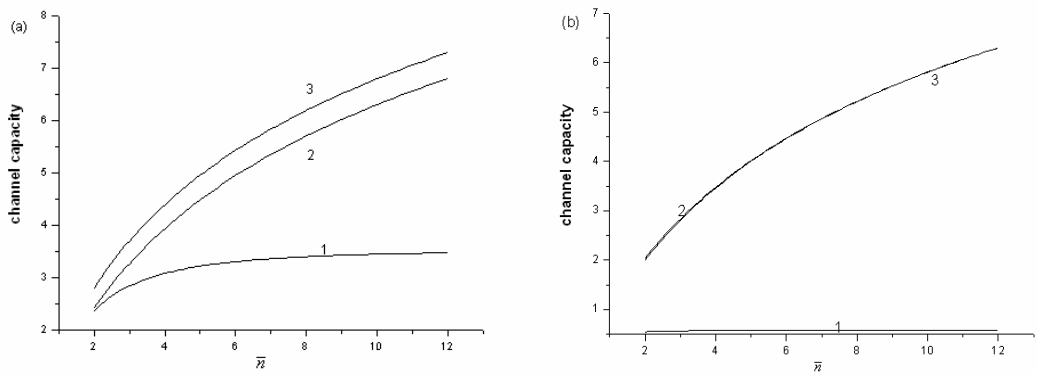

Fig. 5

19